\documentclass[conference]{IEEEtran}

\IEEEoverridecommandlockouts

\usepackage{cite}
\usepackage{multirow}
\usepackage{amsmath,amssymb,amsfonts}
\usepackage{algorithmic}
\usepackage{graphicx}
\usepackage{textcomp}
\usepackage{xcolor}

\usepackage{tikz}
\usetikzlibrary{svg.path}
\usepackage{siunitx}
\usepackage{authblk}
\usepackage{url}

\def\BibTeX{{\rm B\kern-.05em{\sc i\kern-.025em b}\kern-.08em
    T\kern-.1667em\lower.7ex\hbox{E}\kern-.125emX}}

\usepackage{fancyhdr}
\fancypagestyle{firstpage}
{
    \fancyhead[L]{\footnotesize \textcopyright 2026 IEEE.  Personal use of this material is permitted. Permission from IEEE must be obtained for all other uses, in any current or future media, including reprinting/republishing this material for advertising or promotional purposes, creating new collective works, for resale or redistribution to servers or lists, or reuse of any copyrighted component of this work in other works. The paper is accepted at ISQED'26.}
    \fancyhead[R]{}
}

\begin{document}

\title{Cross-Layer Co-Optimized LSTM Accelerator for Real-Time Gait Analysis}


\author{Mohammad Hasan Ahmadilivani}
\author{Levent Aksoy}
\author{Mohammad Eslami}
\author{Jaan Raik}
\author{Alar Kuusik\vspace{-4mm}}

\affil{Tallinn University of Technology, Tallinn, Estonia\vspace{-4mm}}
\affil{\{mohammad.ahmadilivani, levent.aksoy, mohammad.eslami, jaan.raik, alar.kuusik\}@taltech.ee\vspace{-4mm}}

\maketitle
\thispagestyle{firstpage}

\begin{abstract}

Long Short-Term Memory (LSTM) neural networks have penetrated healthcare applications where real-time requirements and edge computing capabilities are essential. Gait analysis that detects abnormal steps to prevent patients from falling is a prominent problem for such applications. Given the extremely stringent design requirements in performance, power dissipation, and area, an Application-Specific Integrated Circuit (ASIC) enables an efficient real-time exploitation of LSTMs for gait analysis, achieving high accuracy. To the best of our knowledge, this work presents the first \mbox{cross-layer} co-optimized LSTM accelerator for real-time gait analysis, targeting an ASIC design. We conduct a comprehensive design space exploration from software down to layout design. We carry out a bit-width optimization at the software level with \mbox{hardware-aware} quantization to reduce the hardware complexity, explore various designs at the \mbox{register-transfer} level, and generate alternative layouts to find efficient realizations of the LSTM accelerator in terms of hardware complexity and accuracy. The physical synthesis results show that, using the $65\;nm$ technology, the die size of the accelerator's layout optimized for the highest accuracy is $0.325\;mm^2$, while the alternative design optimized for hardware complexity with a slightly lower accuracy occupies $15.4\%$ smaller area. Moreover, the designed accelerators achieve accurate gait abnormality detection $4.05\times$ faster than the given application requirement.



\end{abstract}

\begin{IEEEkeywords}
Gait analysis, LSTM accelerator, design space exploration, ASIC design
\end{IEEEkeywords}

\section{Introduction} \label{sec:introduction}

Recent advances in Artificial Intelligence (AI) and Deep Learning (DL) have presented a promising future to achieve effective solutions in medical applications~\cite{manne2021application,rahman2024machine,chakraborty2024machine}. Abnormal gait detection and fall prevention are profound research challenges in gait analysis for patients with neurological disorders, particularly those with neuromuscular conditions \cite{harris2022survey}. These patients often exhibit significant variability and deviations from normal gait patterns, which not only increase their risk of falling~\cite{pirker2017gait}, but also complicate the real-time analysis. The unpredictable and dynamic nature of their gait requires robust and adaptive systems capable of identifying irregularities accurately and preventing falls in real-world scenarios \cite{popovic2014advances}.

Long Short-Term Memory (LSTM) Neural Networks (NNs) have demonstrated strong efficacy in identifying abnormal gait patterns~\cite{khan2024deep,ordonez2016deep,rostovski2024real}. LSTMs incorporate recurrent computations that enable the retention of long-term temporal dependencies, making them particularly suitable for analyzing time-series data~\cite{van2020review}, including gait analysis. The deployment of such models in wearable systems necessitates on-device processing to ensure privacy, low latency, and robustness. 
To meet the stringent requirements of wearable gait analysis applications, such as high performance, low power consumption, and small area footprint, Application-Specific Integrated Circuits (ASICs) represent a promising hardware platform. Their ability to combine high performance with reduced power dissipation and area can facilitate the integration of LSTM-based abnormal gait detection into practical medical devices~\cite{tripathi2022hardware}.

Over the years, ASIC design architectures for LSTM NNs have been proposed for domains such as speech recognition and natural language processing~\cite{conti2018chipmunk,wu2019a3,azari2020elsa,kadetotad2020an8,khan2024digit}. In contrast, \mbox{real-time} gait analysis presents fundamentally different challenges, originating from the distinct characteristics of gait signals and their application context~\cite{miao2021diagnosis}. Unlike speech signals, which are high-frequency and relatively continuous, gait signals are sparse, multi-phased, and strongly influenced by external and biomechanical factors, such as surface hardness, stride length, walking speed, and joint kinematics \cite{wang2024rdgait}. These properties impose stringent requirements on an ASIC design for gait analysis, where the objectives extend beyond accurate pattern recognition to include abnormal gait detection within a single step and robustness to inter- and intra-gait variability. Furthermore, it poses strict design constraints on performance, power dissipation, and chip area to enable real-time processing on resource-limited wearable platforms.

To the best of our knowledge, no LSTM accelerator design exists for real-time abnormality detection for gait analysis in the literature. In this work, we present the first LSTM accelerator, which targets an ASIC design for real-time gait analysis, taking advantage of a cross-layer co-optimized methodology. We carry out a comprehensive design space exploration for bit-width optimization on an \mbox{application-specific} LSTM 
to reduce the hardware complexity and present it as an open-source software tool. 
Furthermore, we design the accelerator using multiple optimization techniques, including parallelism and resource sharing, where the bit-widths of parameters and operations are configured. We introduce two physical layouts of designs with the best hardware complexity and accuracy. The main contributions of this work are as follows:

\begin{figure*}
    \centering
    \parbox{8.9cm}{\centerline{\includegraphics[width=9.8cm]{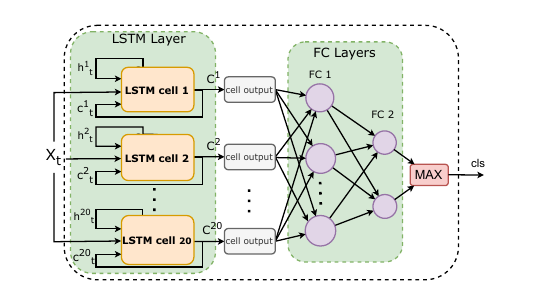}}} \
    \parbox{8.9cm}{\centerline{\includegraphics[width=9.8cm]{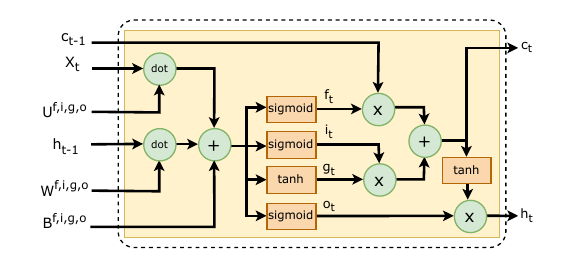}}} \
    \parbox{8.9cm}{\centerline{\scriptsize (a)}}\
    \parbox{8.9cm}{\centerline{\scriptsize (b)}}\
    \caption{(a)~LSTM NN structure; (b)~LSTM cell.}
    \label{fig:lstm}
\end{figure*}

\begin{itemize}
    \item A systematic approach to reduce the hardware complexity, taking into account the accuracy at the software level by exploring the bit-width of parameters and operations of the LSTM through an accurate software simulation corresponding to the accelerator in hardware;
    \item Behavioral design of an LSTM accelerator, where \mbox{bit-widths} of parameters and operations are configured, enabling exploration of alternative implementations with different hardware complexity and accuracy;
    \item Physical designs of two LSTM accelerators, one with the smallest area and the other with the best accuracy, validated on various disease data and pre-trained LSTM models in software. 
\end{itemize}

The physical synthesis results show that the design with the best accuracy has a die size of $0.325\;mm^2$ with a power dissipation of $2.089\;mW$, while the one with the smallest area has $15.4\%$ smaller area and $10.35\%$ lower power dissipation with a tolerable degradation in accuracy with respect to the one with the best accuracy. When compared to the state-of-the-art LSTM accelerators, our design has the smallest area among those under the same technology node.


The rest of this paper is organized as follows: Section~\ref{sec:gait-analysis-lstm} presents the application of LSTM in gait analysis, and Section~\ref{sec:method} introduces the design methodology, both in software and hardware. The experimental setup and results are given in Section \ref{sec:results}, and Section~\ref{sec:conclusions} concludes the paper.

\section{Real-Time Gait Analysis using LSTMs}
\label{sec:gait-analysis-lstm}

The gaits dataset comprises 22 healthy individuals of varying genders, ages, heights, and body weights, used in previous studies ~\cite{rostovski2024real,ahmadilivani2023analysis}. To simulate pathological and various gait patterns, clinical data representing four distinct diseases, i.e., Ataxia, Diplegia, Hemiplegia, and Parkinson’s, are recorded during straight-line walking exercises, conducted under the supervision of a qualified physiotherapist. Each walking step in the dataset is individually labeled as either normal or abnormal. The dataset consists of time-series signals captured by a tri-axial gyroscope, along with the computed magnitude of the recorded motion. To train the NN and enable a detection within a step, each step is augmented into multiple windows with a shifting window, each of which includes $96$ signal samples ($40\%$ of a full step on average), and is considered as an individual input. 

An LSTM NN is developed to distinguish between abnormal and normal steps, separately trained for each disease. The structure of the designed LSTM is shown in Fig.~\ref{fig:lstm}(a). It comprises one LSTM layer with $20$ cells, which is determined as the optimal point in terms of accuracy by an exploration of the number of cells, ranging from $10$ to $30$, through comprehensive experiments. The outputs of the LSTM layer are fed to the Fully-Connected (FC) layer after processing all samples of the time-series inputs in a window of steps. The FC part has two output neurons, representing the normal and abnormal classes. At the output of the LSTM NN, the neuron with the maximum value determines the classification result.


The computations in an LSTM cell are depicted in Fig.~\ref{fig:lstm}(b), where $X_t$ is a time-series input at time $t$, $c^{n}_{t-1}$ and $h^n_{t-1}$ are recurrent inputs, i.e., the outputs of the LSTM cell $n$ at time $t-1$ with $1 \leq n \leq 20$, $U$ and $W$ are the weights of the cell that involve with inputs and recurrent inputs, respectively, and $B$ is the bias parameter. LSTM has four gates in its operations, i.e., \textit{i}, \textit{f}, \textit{g}, and \textit{o}, and each has a different set of $U$, $W$, and $B$ parameters. The LSTM layer processes $96$ data samples with a sampling ratio of $\frac{1}{256}s$. 
After processing these $96$ samples, the output $C^n$ is fed to the FC layer as shown in Fig.~\ref{fig:lstm}(a). 

The FC part has two layers with 20 and 2 neurons, respectively, and each neuron computes as given in Eq.~\ref{eq:fc}:
\begin{equation}
N_i^l = \sum_{j=0}^{m} X_j^{l-1} \times  W_{ij}^l + B_{i}^l
\label{eq:fc}
\end{equation}
where \textit{W} and \textit{B} represent weights and bias, respectively, and \(X^{l-1}\) is $m$ input activations from a previous layer \textit{l-1} with $1 \leq l \leq 2$ and $1 \leq i \leq 20$. Note that the outputs of the first FC layer \textit{FC1} also pass through a ReLU function. As a design requirement, the LSTM accelerator should provide the classification output within $3.9\;ms$ after receiving the last ($96^{th}$) sample data, considering the sampling ratio of the inputs.

Table~\ref{tab:params-count} presents the number of parameters for each component of the LSTM NN separately. Note that the LSTM NN has a total of $2462$ parameters. Moreover, Table~\ref{tab:full-precision-acc} presents the achieved accuracy and F1-score for the presented LSTM NN on each anomaly separately, which is trained in full-precision and ensured not to overfit during the training. 

\begin{table*}[t!]
\centering
\caption{Number of parameters in all layers of the LSTM NN.}
\label{tab:params-count}
\footnotesize
\begin{tabular}{|l|c|c|c|c|c|c|c|}
\hline
Parameters   & $U^{f,i,g,o}$    & $W^{f,i,g,o}$   & $B^{f,i,g,o}$  & $W^{FC1}$   & $B^{FC1}$  & $W^{FC2}$  & $B^{FC2}$  \\ \hline \hline
\# of cells/neurons & 20   & 20  & 20 & 20  & 20 & 2 & 2  \\ \hline
\# of gates         & 4    & 4   & 4  & -   & -  & -  & - \\ \hline
\# of parameters in a gate/neuron & 20   & 4   & 1  & 20  & 1  & 20  & 1  \\ \hline
Total               & 1600 & 320 & 80 & 400 & 20 & 40 & 2  \\ \hline
\end{tabular}%
\end{table*}

\begin{table}[t]
\centering
\caption{Accuracy and F1-score of the LSTM NN.}
\footnotesize
\begin{tabular}{|l|c|c|}
\hline 
Metrics      & Accuracy & F1-score \\ \hline \hline
Ataxia       & 87.53\%  & 72.28\%  \\ \hline
Diplegia     & 81.48\%  & 74.74\%  \\ \hline
Hemiplegia   & 87.11\%  & 67.47\%  \\ \hline
Parkinson's  & 82.08\%  & 72.50\%  \\ \hline
\end{tabular}%
\label{tab:full-precision-acc}
\end{table}

\section{Cross-Layer Co-Optimized LSTM Accelerator}
\label{sec:method}

\begin{figure}[t]
\centering
    \includegraphics[width=0.4\textwidth]{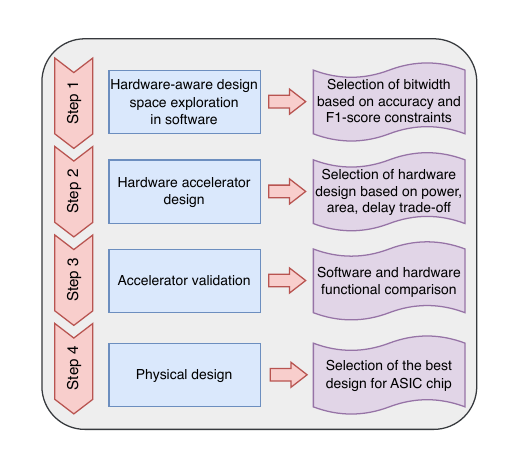}
    \caption{Steps in the design of an LSTM NN hardware accelerator.}
    \label{fig:method}
\end{figure}

Fig.~\ref{fig:method} shows the main steps in the design of an efficient LSTM NN accelerator in terms of hardware complexity and accuracy, which can be generalized for any LSTM NNs. The following sections describe each step in detail. 

\subsection{Hardware-Aware Design Space Exploration in Software} 

The first step focuses on the bit-width optimization of the LSTM NN in software to find a hardware-efficient design. The objective is to minimize the bit-width of parameters and operations while meeting the accuracy and F1-score constraint of the application. Note that the application requires less than $1\%$ degradation in accuracy and F1-score of the optimized NNs compared to their full-precision outputs given in Table~\ref{tab:full-precision-acc}. We conduct the following procedures, concerning the effect of fixed-point quantization of parameters and operations on accuracy and F1-score while reducing the hardware complexity. 


\subsubsection{Detailed Design of LSTM NN} We implement a detailed software description of the LSTM NN. For the fixed-point quantization of computations, it is necessary to implement the LSTM NN down to atomic operations that correspond to hardware units, including multiplication and addition. The nonlinear activation functions, i.e., sigmoid, tanh, and ReLU, the LSTM cells, and all FC layers are also implemented. 

\subsubsection{Fixed-point Quantization of the LSTM NN} We quantize all the full-precision parameters and input/output values of computations to fixed-point values. A value $x$ in full-precision $x_{FP}$ can be quantized into a fixed-point value $x_{FxP}$ in data format of $FxP(b, f)$ with total bits of $b$ and fractional bits of $f$ as given in  Eqs.~\eqref{eq:fxp1} and~\eqref{eq:fxp2}: 
\begin{gather}
    y = sign(x_{FP}) \times \frac{|x_{FP}| + \epsilon}{2^{-f}} \times {2^{-f}}; \quad \epsilon = 2^{-f} \label{eq:fxp1} \\
    x_{FxP} = 
    \begin{cases}
        Max(FxP(b,f))&y > Max(FxP(b,f)) \\
        Min(FxP(b,f))&y < Min(FxP(b,f)) \\
        y&otherwise
    \end{cases}
     \label{eq:fxp2}
\end{gather}
where Eq.~\eqref{eq:fxp1} quantizes the fractional part of a full-precision value, while $\epsilon$ represents the rounding error and Eq.~\eqref{eq:fxp2} clamps the value obtained in  Eq.~\eqref{eq:fxp1} to the minimum and maximum values that can be represented by $FxP(b, f)$.

To quantize the detailed operations of LSTM NN, the operations' inputs, i.e., weights, bias, and data, and outputs are quantized to a $FxP(b,f)$ format using Eqs.~\eqref{eq:fxp1} and~\eqref{eq:fxp2}. Input time-series data are always quantized into $FxP(10,8)$. Since the design of activation functions \textit{sigmoid} and \textit{tanh} in hardware is highly complex, we apply a polynomial approximation~\cite{fonseca2016fpga, li18} as follows:

\begin{gather}
    \footnotesize
    sigmoid(x) = 
    \begin{cases}
        0                                & \phantom{-6 <\;} x \leq -6 \\
        0.00642x^2 + 0.07176x + 0.20323  & -6 < x \leq -3 \\
        0.04059x^2 + 0.27269x + 0.50195  & -3 < x \leq 0  \\
        -0.04058x^2 + 0.27266x + 0.49805 & \phantom{-} 0 < x \leq 3  \\
        -0.00642x^2 + 0.07175x + 0.79675 & \phantom{-} 3 < x \leq 6  \\
        1                                & \phantom{-6 <\;} x > 6
    \end{cases}
    \nonumber
\end{gather}
\begin{gather}
    \footnotesize
    tanh(x) = 
    \begin{cases}
        -1                                 & \phantom{-6 <\;} x \leq -3\\
        0.09007x^2 + 0.46527x - 0.39814    & -3 < x \leq -1\\
        0.31592x^2 + 1.08381x + 0.00314    & -1 < x \leq 0 \\
        - 0.31676x^2 + 1.08538x - 0.00349  & \phantom{-} 0 < x \leq 1 \\
        - 0.09013x^2 + 0.46509x + 0.39878  & \phantom{-} 1 < x \leq 3 \\
        1                                  & \phantom{-6 <\;} x > 3
    \end{cases}
    \nonumber
\end{gather}
where all coefficients and operations are quantized into $FxP(18,13)$ to keep a high computational precision.

\subsubsection{Design Space Exploration for Bit-width Optimization} To minimize the bit-width of the parameters and operations without compromising the accuracy, a comprehensive design space exploration is conducted. Various bit-widths are applied to parameters and operations of the LSTM NNs. In hardware, the bit-width of parameters affects the memory size, and the \mbox{bit-width} of parameters and operations has an impact on the size of the data path and arithmetic logic. We pinpoint the computations in the software to the corresponding hardware blocks, so the quantization applied at the software level mimics its impact on the functionality of the LSTM NN in hardware. 

We obtain the accuracy and \mbox{F1-score} of the LSTM NN at the software level, exploring possible bit-width configurations. Considering the application constraint in the tolerable degradation in accuracy and \mbox{F1-score}, i.e., $<1\%$, we select the bit-width configurations that lead to designs with small hardware complexity. For the sake of generality, we apply the exploration to all four LSTM NNs given in Table~\ref{tab:full-precision-acc} and consider the \mbox{worst-case} accuracy and F1-score degradation among all of them. 

\begin{figure}[t]
    \centerline{\includegraphics[width=9.0cm]{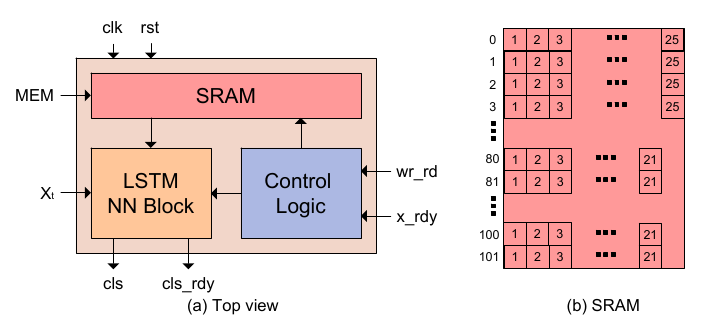}}
	\caption{LSTM accelerator design in hardware.}
	\label{fig:lstm_top}
\end{figure}

\subsection{LSTM NN Accelerator Design in Hardware} 

The top-level view of the LSTM NN accelerator is given in Fig.~\ref{fig:lstm_top}(a). The input \textit{MEM} denotes all the signals related to the SRAM, such as memory address, data, and enable, and the input $X_t$ stands for the data to be processed by the LSTM NN block. The input \textit{wr\_rd} denotes the write or read mode, where the parameters are written to the SRAM during the initialization or read from the SRAM during the execution, respectively, and \textit{x\_rdy} indicates that the LSTM input $X_t$ is ready for execution. While the output \textit{cls} is the classification result, \textit{cls\_rdy} indicates that the classification output is ready. 

To avoid the data transfer between the off-chip memory and accelerator, a design with an on-chip memory~\cite{chen14} is preferred. The SRAM stores the parameters of LSTM cells and FC layers. 
To reduce the number of SRAM accesses, its contents are organized as shown in Fig.~\ref{fig:lstm_top}(b). As given in Table~\ref{tab:params-count}, all $25$, $(20+4+1)$, parameters related to each gate, i.e., $i$, $f$, $g$, and $o$, in LSTM cells are placed in a consecutive order between addresses $0$ and $79$. All $21$, $(20+1)$, parameters of the first FC layer are placed between addresses $80$ and $99$, and the $21$, $(20+1)$, parameters of the second FC layer are placed between addresses $100$ and $101$. In our experiments, three FxP configurations, i.e., (10,8), (9,7), and (8,6), leading to $24620$, $22158$, and $19696$ bits, are considered for parameters to explore the trade-off between hardware complexity and accuracy. Since the memory compiler cannot generate a single SRAM for these three configurations, two SRAM banks with the same number of addresses and data sizes are used.

The LSTM NN block initially makes computations in parallel related to $i$, $f$, $g$, and $o$ gates in each LSTM cell, then in each neuron of the first FC layer and in each neuron of the second FC layer, generating the classification result, as shown in Fig.~\ref{fig:lstm}. All the common operations, e.g., those that realize the dot product and activation functions, are shared during computation. Whenever the outputs of the LSTM cell, i.e., $c_t$ and $h_t$, are generated for one data sample, they are stored for the next computation. To balance the trade-off between hardware complexity and accuracy, while the size of all multiplication operations, including the ones in the dot product and activation functions, is fixed to the given $FxP$ data format, the size of all addition operations is not restricted. 

The control logic synchronizes the processes of writing/reading the parameters to/from the SRAM based on the \textit{wr\_rd} signal. The execution of the LSTM NN block is controlled by a \textit{counter} since the parameters, which are required for the computation at a specific time, are known beforehand. The total number of clock cycles required to compute the classification output is given as $96 \times 20 \times (4+1) + (20+1) + (2+1)$, equal to $9624$. In the first product, $96$ is the number of data samples mentioned in Section~\ref{sec:gait-analysis-lstm}, $20$ is the number of LSTM cells, and $(4+1)$ denotes the computation for four gates $i$, $f$, $g$, and $o$ and one clock cycle to store the computation. The second and third sums $(20+1)$ and $(2+1)$ denote the number of neurons in the first and second FC layers, respectively, and one clock cycle to store the computation. After the classification output is obtained, the whole process restarts.

The accelerator is described at register-transfer level (RTL) in Verilog, where the bit-width of parameters and operations is made configurable to explore the trade-off between hardware complexity and accuracy.

\subsection{Accelerator Validation}

After designing the accelerator at RTL, it is validated through a comprehensive simulation with the software implementation, where the objective is to observe whether the outputs of the hardware accelerator match the functionality of the LSTM NN in software. Initially, the dot product and activation functions in the LSTM cell, and the computation of neurons in FC layers are validated. In this regard, their inputs are exported from the software and simulated in hardware. Then, the outputs of the hardware design are compared with those of the software implementation. Finally, the whole accelerator in RTL is validated through simulation, and the accuracy and F1-score values are obtained based on the same inputs in software and hardware.  

\subsection{Physical Design} 
Based on the evaluations in the design and validation of accelerators with various configurations on the \mbox{bit-width} of parameters and operations, two designs, one with the smallest area and the other with the best accuracy, are selected for physical design. After the logic synthesis on the RTL design, physical synthesis is performed using the same technology library and Process Design Kit (PDK) to ensure consistency across both designs. The SRAM is instantiated as a \mbox{black-box} component in the gate-level netlist since it is compiled independently and is integrated into the overall design and placed alongside other logic elements in the physical design.

\section{Experimental Results} 
\label{sec:results}

This section initially presents the cross-layer optimization results of the proposed methodology, and then the gate-level synthesis results of various accelerator designs with different design parameters. Finally, it introduces the physical synthesis results of selected accelerators.

\subsection{Hardware-Aware Bit-Width Exploration}
\label{subsec:bit-exploration}

The software description of the LSTM NN is implemented in Python with the use of the PyTorch library. All the detailed operations of the LSTM NN are implemented to enable a hardware-aware quantization, which is presented as an \mbox{open-source} tool\footnote{\url{https://github.com/mhahmadilivany/LSTM-ASIC-optimization}}. 

\begin{figure}[t]
\centering
    \includegraphics[width=0.50\textwidth]{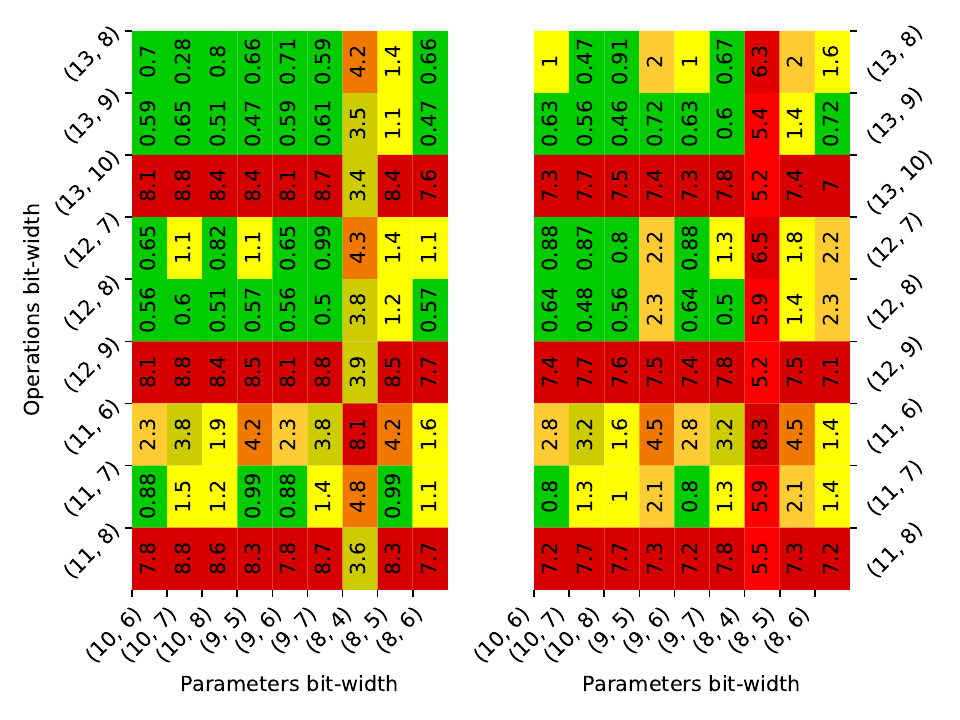}
    \caption{Exploration of degradation in accuracy (left) and F1-score (right) over different bit-widths of parameters and operations. Green color represents lower degradation, yellow, orange, and red colors represent higher degradation, respectively.  }
    \label{fig:acc-f1-exploration}
\end{figure} 

To perform the hardware-aware bit-width design space exploration, we quantize the parameters and operation values separately. We profile their values for the design space exploration in a way that the bit-widths lead to a minimal overflow during computations. The full inference of data is executed to obtain accuracy and F1-score under various bit-widths for each dataset separately. Fig.~\ref{fig:acc-f1-exploration} shows the worst-case accuracy and F1-score degradation across all four LSTM NNs with respect to those given in Table~\ref{tab:full-precision-acc}. The values marked by green show the bit-width configurations where the accuracy/F1-score drop is below $1\%$, and the other colors represent higher degradations. 


\begin{table}[t]
\centering
\caption{Selected fixed-point configurations.}
\footnotesize
\label{tab:selected-confs}
\begin{tabular}{|@{\hskip4pt}l@{\hskip4pt}|@{\hskip4pt}c@{\hskip4pt}|@{\hskip4pt}c@{\hskip4pt}|@{\hskip4pt}c@{\hskip4pt}|@{\hskip4pt}c@{\hskip4pt}|@{\hskip4pt}c@{\hskip4pt}|@{\hskip4pt}c@{\hskip4pt}|@{\hskip4pt}c@{\hskip4pt}|}
\hline
Configuration                                                   & \#1       & \#2       & \#3       & \#4       & \#5       & \#6       & \#7       \\ \hline \hline
\begin{tabular}[c]{@{}c@{}}Parameter\\ bit-width\end{tabular}  & (10,8) & (10,8) & (10,8) & (9,7)  & (9,7)  & (9,7)  & (8,6)  \\ \hline
\begin{tabular}[c]{@{}c@{}}Operation\\ bit-width\end{tabular} & (13,8) & (13,9) & (12,8) & (13,8) & (13,9) & (12,8) & (13,9) \\ \hline
\end{tabular}%
\end{table}

\begin{table}[t]
    \centering
    \caption{Gate-level synthesis results of accelerator designs.}
    \footnotesize
	\begin{tabular}{|c|c|c|c|c|c|c|}
		\hline
		Parameter & Operation &  Config. & \begin{tabular}[c]{@{}c@{}}Area \\ (${\mu m}^2$)\end{tabular} & \begin{tabular}[c]{@{}c@{}}Delay\\ ($ns$)\end{tabular} & \begin{tabular}[c]{@{}c@{}}Power\\ ($nW$)\end{tabular} \\ 
		\hline \hline
		\multirow{3}{*}{FxP(10,8)} & FxP(13,8) & \#1 & 104633 & 15.6 & 720963 \\
                                           & FxP(13,9) & \#2 & 104487 & 14.7 & 722755 \\
                                           & FxP(12,8) & \#3 & 96345  & 14.5 & 686553 \\
        \hline                                                        
		\multirow{3}{*}{FxP(9,7)}  & FxP(13,8) & \#4 & 100283 & 15.5 & 670316 \\
                                           & FxP(13,9) & \#5 & 100153 & 15.1 & 662930 \\
                                           & FxP(12,8) & \#6 & 92152  & 14.6 & 474603 \\
        \hline                                                        
		FxP(8,6)                   & FxP(13,9) & \#7 & 89996  & 15.2 & 659818 \\
        \hline
	\end{tabular}
    \label{tab:glsr}
\end{table}

\begin{table}[t]
    \centering
    \caption{Gate-level synthesis results of the design with the \textit{configuration~\#7} under strict delay constraints.}
    \footnotesize
	\begin{tabular}{|c|c|c|c|}
		\hline
		Design & Area (${\mu m}^2$) & Delay ($ns$) & Power ($nW$) \\ 
		\hline \hline
          1 & 89996  & 15.2 & 659818  \\
          2 & 93161  & 7.4  & 3330029 \\
          3 & 93696  & 6.9  & 3604827 \\
          4 & 95448  & 6.4  & 3954104 \\ 
          5 & 98255  & 5.9  & 4649098 \\
          6 & 100113 & 5.4  & 5328803 \\
          7 & 105524 & 4.9  & 5758332 \\
        \hline
	\end{tabular}
    \label{tab:delayopt}
\end{table}

Notably, it is observed from Fig.~\ref{fig:acc-f1-exploration} that having sufficient bits for both integer and fraction bits is necessary to achieve a small error. On one hand, a low number of bits for integers results in accuracy degradation, since large values cannot be represented, as in $FxP(13, 10)$, $FxP(12, 9)$, and $FxP(11, 8)$ for operations, in which 2 bits plus one sign bit are for integers. On the other hand, a small number of fractions reduces the precision of computations, as in $FxP(8, 4)$ for parameters, in which fractions are represented by 4 bits.

Moreover, according to the \mbox{gate-level} synthesis results, a large number of fraction bits results in a small hardware complexity for multipliers. We synthesized fixed-point multipliers with two inputs and different bit-width configurations and observed that the designs with a large fraction value $f$ while the total number of bits $b$ is the same, have slightly less hardware complexity when compared to those with a small fraction value. Based on these facts,  $7$ bit-width configurations are selected from those presented in Fig.~\ref{fig:acc-f1-exploration} for hardware-efficient design exploration, where both accuracy and \mbox{F1-score} have an error of less than $1\%$ compared to the \mbox{full-precision} execution. Table~\ref{tab:selected-confs} presents these configurations.

\subsection{Gate-Level Synthesis}
\label{subsec:glsr}

Table~\ref{tab:glsr} introduces the gate-level synthesis results of accelerators under the bit-width configurations given in Table~\ref{tab:selected-confs}, 
where \textit{area} is the total area, \textit{delay} is the delay in the critical path, and \textit{power} is the total power dissipation, which are obtained after logic synthesis by Cadence Genus using a commercial $65\;nm$ gate library with the aim of area optimization. Note that the SRAM, which stores the parameters, is excluded from these designs. 

Observe from Table~\ref{tab:glsr} that the bit-width of parameters has a significant impact on the accelerator area, e.g., when it is $8$, the area is respectively $10.1\%$ and $13.8\%$ less than that of designs when it is $9$ and $10$ under the configuration of $FxP(13,9)$ for the operation bit-width. Also, the bit-width of operations has a significant impact on the accelerator area, e.g., when it is $12$,  the area is respectively $7.79\%$ and $7.92\%$ less than that of the designs when it is configured as $FxP(13,9)$ and $FxP(13,8)$ while the parameter bit-width is $10$. Similar trends are observed on the accelerator designs while the parameter bit-width is $9$. Note also that different configurations on the same operation bit-width, e.g., $FxP(13,9)$ and $FxP(13,8)$, have a slight impact on the area of accelerator designs. Similar to the area, the bit-width of parameters and operations has a significant impact on power dissipation.

Based on the gate-level synthesis results given in Table~\ref{tab:glsr}, the \textit{configuration \#7} with $FxP(8,6)$ and $FxP(13,9)$ for \mbox{bit-width} of parameters and operations, respectively, is chosen for the layout design since it leads to the least complex design. 

To further explore the tradeoff between the area and delay, the design with the \textit{configuration \#7} is synthesized under strict delay constraints used to find the maximum frequency that the design can support. The gate-level results of these designs are given in Table~\ref{tab:delayopt}, where the first design is obtained with a relaxed delay constraint as given in Table~\ref{tab:glsr}.
Observe from Table~\ref{tab:delayopt} that the delay of the original accelerator can be reduced $3.1\times$, but increasing the area and power dissipation by $1.17\times$ and $8.72\times$, respectively, indicating that the performance of the design can be further optimized.

\subsection{Accelerator Validation} \label{subsec:validation}

Table~\ref{tab:validation-error} presents the validation results of the computational components between the LSTM NN accelerator and its software description for the \textit{configuration \#4} as an example. Observe that the difference in the output values between the hardware and software simulation is negligible in all components, i.e., it is always less than $2^{-6}$. The average error for $C$ and $H$ outputs is slightly higher than $2^{-6}$ since they accumulate errors through $96$ iterations in the LSTM layer. Nonetheless, such a small error may have an impact on the overall accuracy due to cumulative rounding errors during computations. 

\begin{table}[t]
\centering
\caption{Error in the components implemented in hardware compared to their software implementations for configuration \#4.}
\footnotesize
\label{tab:validation-error}
\begin{tabular}{|l|c|c|}
\hline
Target output & Max error  & Average error \\ \hline \hline
i, f, g, o gates     & 0.01562   & 0.00143      \\ \hline
tanh, sigmoid         & 0.00390   & 0.00025      \\ \hline
C, H                  & 0.05078   & 0.01566       \\ \hline
Neurons in FC            & 0.01562   & 0.00104      \\ \hline
NN full simulation       & 0.03125   & 0.01249       \\ \hline
\end{tabular}%
\end{table}

\begin{table*}[t]
\centering
\caption{Worst-case accuracy and F1-score degradation in hardware across all datasets for different bit-width configurations. }
\footnotesize
\label{tab:full-validation}
\begin{tabular}{|c|c|c|c|c|}
\hline
\#Configuration &
  \begin{tabular}[c]{@{}c@{}}Worst accuracy\\ degradation \\ from software\end{tabular} &
  \begin{tabular}[c]{@{}c@{}}Worst F1-score \\ degradation\\ from software\end{tabular} &
  \begin{tabular}[c]{@{}c@{}}Worst accuracy\\ degradation from\\ full-precision\end{tabular} &
  \begin{tabular}[c]{@{}c@{}}Worst F1-score \\ degradation from\\ full-precision\end{tabular} \\ \hline \hline
\#1 & 0.41\%  & 0.49\%  & 0.89\% & 1.34\% \\ \hline
\#2 & 0.18\%  & 0.05\%  & 1.01\% & 1.15\% \\ \hline
\#3 & 0.11\%  & 0.14\%  & 0.80\% & 1.28\% \\ \hline
\#4 & 0.25\%  & 0.22\%  & 0.53\% & 0.71\% \\ \hline
\#5 & -0.03\% & 0.13\%  & 0.50\%  & 0.49\% \\ \hline
\#6 & 0.06\%  & 0.001\% & 0.50\%  & 0.72\% \\ \hline
\#7 & 0.38\%  & 0.61\%  & 0.91\% & 1.08\% \\ \hline
\end{tabular}%
\end{table*}

Table~\ref{tab:full-validation} presents the worst-case accuracy and F1-score degradation obtained in the hardware accelerator when compared to its software implementations obtained through all datasets in each bit-width configuration. Observe that the overall error in the hardware accelerator designs is generally small compared to the software implementations, although it is more pronounced in some configurations, such as the \textit{configuration \#1 and configuration \#7}. Based on the results given in Table~\ref{tab:full-validation}, the \textit{configuration \#5} with  $FxP(9,7)$ and $FxP(13,9)$ for the bit-width of parameter and operation, respectively, is selected for the layout design since it leads to the least degradation in accuracy and F1-score. 

\subsection{Physical Synthesis}
\label{subsec:pdr}

To gain accurate insights into the practical implications of our architectural choices, such as area utilization, routing complexity, and power dissipation, we proceed with a full physical synthesis of two designs, i.e., the \textit{configuration \#7} with the least hardware complexity and the \textit{configuration \#5} with the best accuracy.  Note that performing physical synthesis enables realistic evaluation beyond RTL metrics, especially in the context of memory integration and floorplanning.

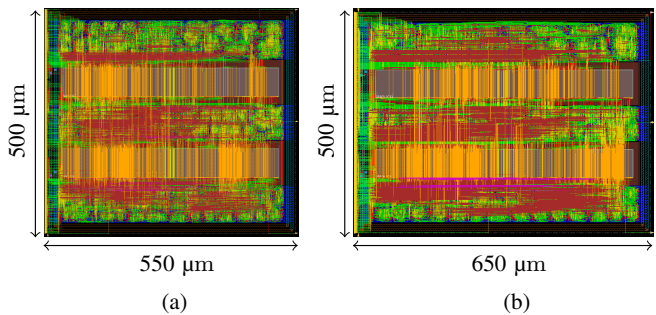
\begin{figure}[t]
    \centering
    \scalebox{1.09}{
    \begin{tikzpicture}
    \node[anchor=south west,inner sep=0] (image1) at (0,0) {\includegraphics[width=0.35\columnwidth, height=0.315\columnwidth]{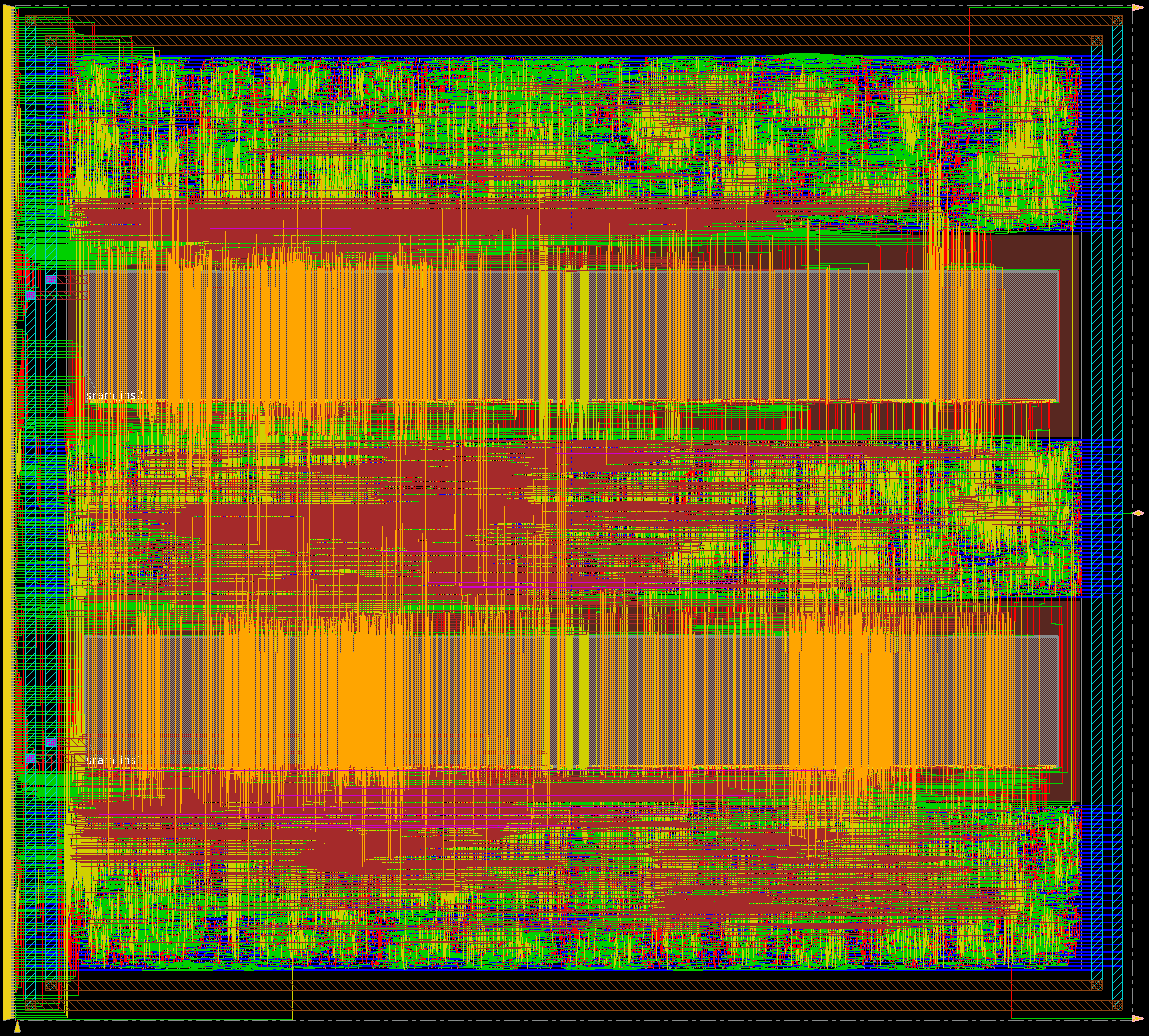}};
        \draw [<->](-.10 , 0) -- (-.10 , 2.77) node[left, font=\footnotesize, rotate=90] at (-.3,2) {\: 500 \si{\micro\meter}};
        \draw [<->](0 , -.10) -- (3.1 , -.10) node[midway, below, font=\footnotesize] {\: 550 \si{\micro\meter}};
        \node[font=\footnotesize] at (1.6,-0.8) {(a)};
    \end{tikzpicture}
    \begin{tikzpicture}
    \node[anchor=south west,inner sep=0] (image1) at (0,0) {\includegraphics[width=0.413\columnwidth, height=0.315\columnwidth]{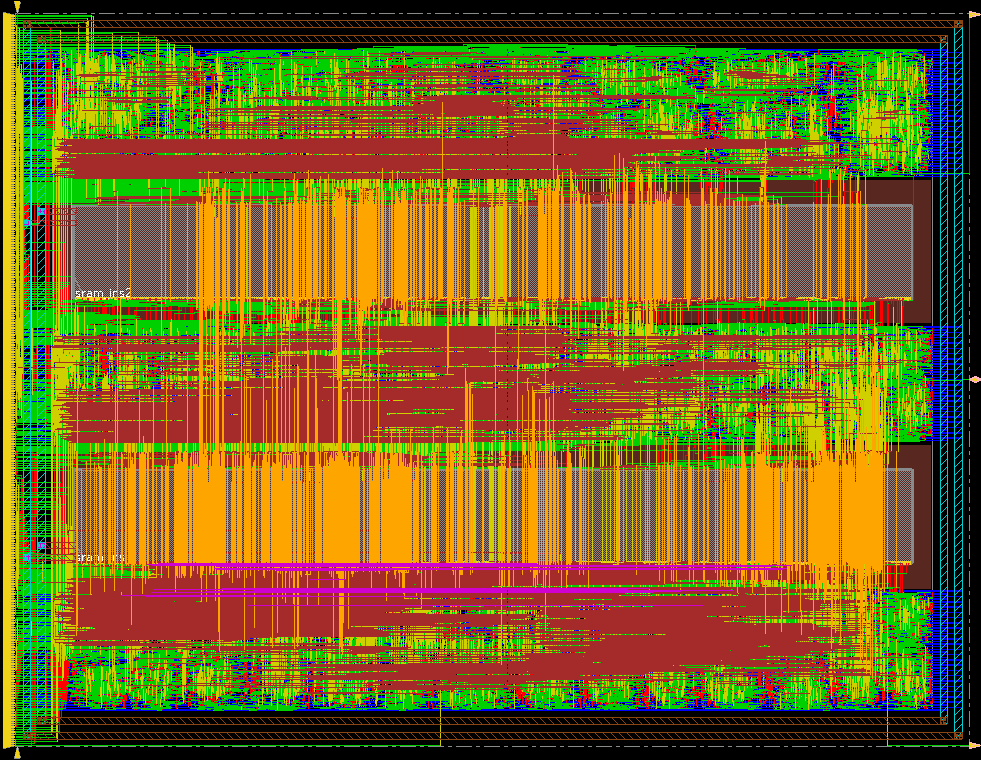} };
        \draw [<->](-.10 , 0) -- (-.10 , 2.77) node[left, font=\footnotesize, rotate=90] at (-.3,2) {\: 500 \si{\micro\meter}};
        \draw [<->](0 , -.10) -- (3.65 , -.10) node[midway, below, font=\footnotesize] {\: 650 \si{\micro\meter}};
        \node[font=\footnotesize] at (2.0,-0.8) {(b)};
    \end{tikzpicture}  
    }
    \caption{Layouts with routed view of the selected designs: (a)~configuration~\#7; (b)~configuration~\#5.}
    \label{fig:layout_routed}
\end{figure}

\begin{figure}[t]
    \centering
    \scalebox{1.14}{
    \begin{tikzpicture}
    \node[anchor=south west,inner sep=0] (image1) at (0,0) {\includegraphics[width=0.32\columnwidth, height=0.288\columnwidth]{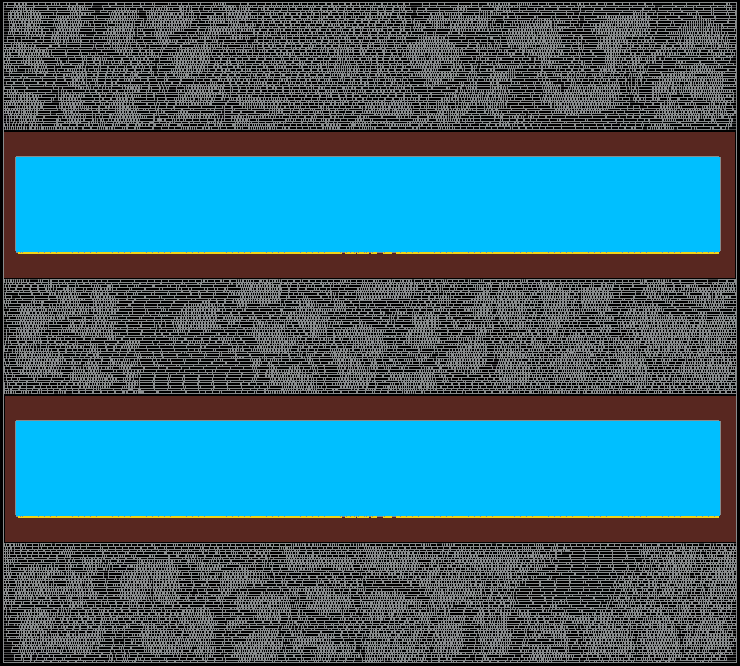}};
        \draw [<->](-.10 , 0) -- (-.10 , 2.53) node[left, font=\footnotesize, rotate=90] at (-.3,1.9) {\: 450 \si{\micro\meter}};
        \draw [<->](0 , -.10) -- (2.82 , -.10) node[midway, below, font=\footnotesize] {\: 500 \si{\micro\meter}};
        \node[font=\footnotesize] at (1.6,-0.8) {(a)};
    \end{tikzpicture}
    \begin{tikzpicture}
    \node[anchor=south west,inner sep=0] (image1) at (0,0) {\includegraphics[width=0.384\columnwidth, height=0.288\columnwidth]{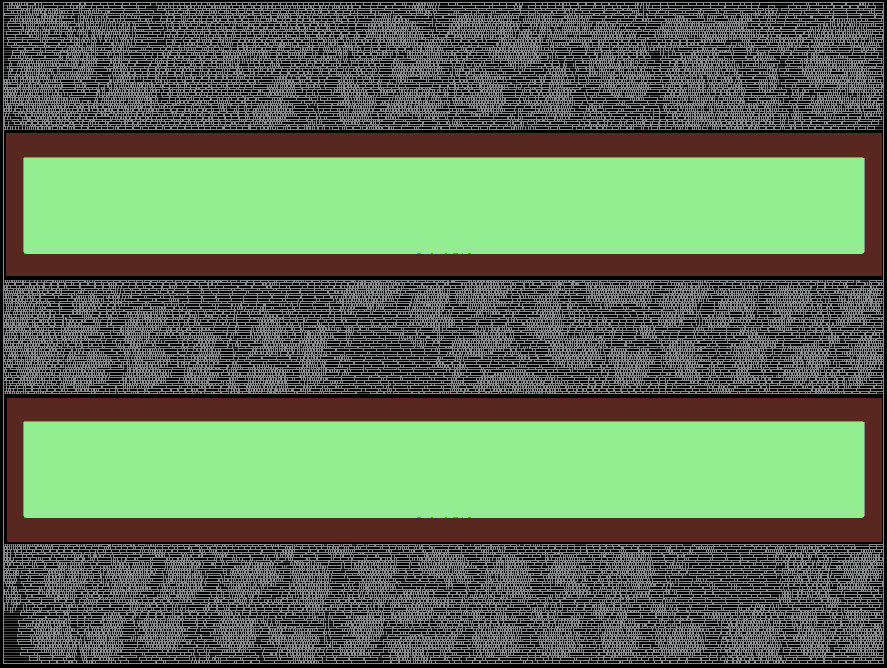}};
        \draw [<->](-.10 , 0) -- (-.10 , 2.53) node[left, font=\footnotesize, rotate=90] at (-.3,1.9) {\: 450 \si{\micro\meter}};
        \draw [<->](0 , -.10) -- (3.4 , -.10) node[midway, below, font=\footnotesize] {\: 600 \si{\micro\meter}};
        \node[font=\footnotesize] at (1.8,-0.8) {(b)};
    \end{tikzpicture}
    }
    \caption{Layouts with placement view of the selected designs: (a)~configuration~\#7; (b)~configuration~\#5.}
    \label{fig:layout_placed}
\end{figure}
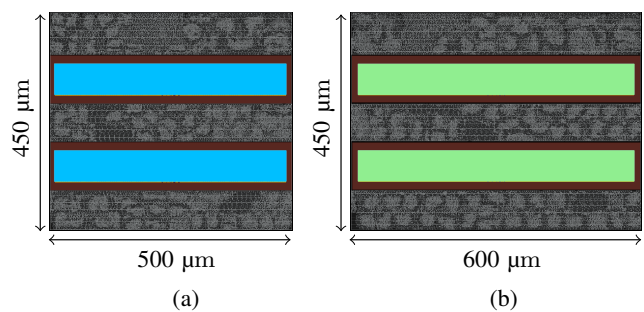

For the physical synthesis of these designs, we employ Cadence Innovus using a commercial $65\;nm$ technology node. Fig.~\ref{fig:layout_routed} illustrates the final layout views after routing. Observe that the \textit{configuration \#5} incorporates a wider SRAM block, resulting in a correspondingly wider die when compared to the \textit{configuration \#7}. Fig.~\ref{fig:layout_placed} presents the layout views after the placement stage. To provide a clear visual comparison of the memory and logic area distribution, the SRAM blocks are highlighted. Observe that the SRAMs in both designs occupy a substantial portion of the core area, nearly comparable to the area consumed by the remaining logic. 

This highlights the importance of careful placement and orientation of the memory blocks to ensure design efficiency. For instance, placing the SRAMs in a vertical orientation or positioning them near the corners of the core area can introduce significant overheads in terms of routing congestion, increased wirelength, and suboptimal power distribution. Such placements may also hinder timing closure due to increased distance from critical components. Therefore, strategic placement of SRAM blocks near the center of the core, with horizontal orientation and proximity to key logic clusters, is crucial for minimizing routing complexity and optimizing overall design performance.

Table~\ref{tab:ppa_layout} summarizes the power, performance, and area results obtained after the physical synthesis stage. It reports key metrics for each design, including \textit{total area}, which refers to the standard cell area only and does not represent the full floorplan area, the \textit{slack}, which denotes the difference between the required arrival time set to $100\;ns$ and delay in the critical path, and various power dissipation values. To provide a clear perspective on the efficiency of the \textit{configuration~\#7} relative to the \textit{configuration~\#5}, the gain in percentage between these two designs is also given. Note that both designs achieve positive slack, indicating successful timing closure under the specified delay constraint. Therefore, while slack values are included for completeness, they are not used as a basis for direct performance comparison, since a higher or lower slack within the positive range does not imply one design operates faster than the other in this context.

\begin{table}[t]
  \centering
  \caption{Physical synthesis results.}
  \footnotesize
  \begin{tabular}{|@{\hskip3pt}l@{\hskip3pt}|@{\hskip3pt}c@{\hskip3pt}|@{\hskip3pt}c@{\hskip3pt}|@{\hskip3pt}c@{\hskip3pt}|}
    \hline
     Design & configuration \#7 & configuration \#5 & Gain \\ 
    \hline \hline
    Total area (\si{\micro\meter\squared})  & 152,369 & 174,537 & 12.70\% \\
    \hline
    Slack (\si{\nano\second})               & 32.224 & 31.372 & NA\\
    \hline
    Internal Power (\si{\milli\watt})       & 1.233 & 1.372 & 10.13\%\\
    \hline
    Switching Power (\si{\milli\watt})      & 0.588 & 0.659 & 10.77\%\\
    \hline
    Leakage Power (\si{\milli\watt})        & 0.006 & 0.007 & 14.27\%\\
    \hline
    Total Power (\si{\milli\watt})          & 1.827 & 2.038 & 10.35\%\\
    \hline
  \end{tabular}
  \label{tab:ppa_layout}
\end{table}

The accelerator design under \textit{configuration \#5} meets the accuracy constraints, and its layout achieves a die size of $0.325\;mm^2$ (Fig.~\ref{fig:layout_routed}(b)) and a total power consumption of $2.038\;mW$. By tolerating the smallest accuracy degradation less than the application constraint, the \textit{configuration \#7} leads to a layout with $15.4\%$ smaller die size and $10.35\%$ lower power consumption. It is important to note that both designs meet the real-time constraints of the application by operating at the frequency of $10\;MHz$. Considering that a time-series gait data with $96$ samples is processed in $9,624$ clock cycles, the classification result is provided in $0.9624\;ms$, which is $4.05\times$ less than the required time, i.e., $3.9\;ms$.

Table \ref{table:sota-comparison} compares our design with the smallest hardware complexity, i.e., \textit{configuration \#7}, with the state-of-the-art LSTM accelerators targeting different applications. Observe that our design achieves $0.8$ TOPS/W energy efficiency and $9.6$ GOPS/mm$^2$ area efficiency and occupies the smallest area when compared to those under the same technology node. Since the energy and area efficiency are influenced by the frequency and the target application possesses a low frequency, the energy and area efficiency in our ASIC design are lower than those in other state-of-the-art accelerators.

\begin{table*}[hbt!]
\centering
\caption{Comparison with the state-of-the-art LSTM NNs.}
\footnotesize
\label{tab:my-table}
\begin{tabular}{|l|c|c|c|c|c|}
\hline
Design                           & Ours & \cite{conti2018chipmunk} & \cite{wu2019a3} & \cite{azari2020elsa} & \cite{kadetotad2020an8} \\ \hline \hline 
Application     &   \begin{tabular}[c]{@{}c@{}}Gait \\ analysis\end{tabular}     &   \begin{tabular}[c]{@{}c@{}}Speech \\ recognition\end{tabular}  &   \begin{tabular}[c]{@{}c@{}}Language \\ processing\end{tabular}   &  \begin{tabular}[c]{@{}c@{}}Language \\ modeling \end{tabular}  &  \begin{tabular}[c]{@{}c@{}}Speech \\ recognition\end{tabular}           \\ \hline
Technology (nm)               & 65        & 65       & 40       & 65       & 65       \\ \hline
Area (mm$^2$)                 & 0.152     & 1.57     & 0.45     & 2.62     & 7.74  \\ \hline
Power (mW)                    & 1.827     & 1.24     & 6.16     & 20.4     & 67.3     \\ \hline
On-Chip Memory (KB)           & 2.704     & 82       & 88.5     & 106      & 297    \\ \hline
Voltage (V)                   & 1.2       & 0.75     & 1.1      & 1.1      & 1.1      \\ \hline
Frequency (MHz)               & 10        & 20       & 200      & 322      & 80       \\ \hline 
Energy Efficiency (TOPS/W)    & 0.8       & 3.08     & 3.89     & 1.32     & 2.45     \\ \hline
Area Efficiency (GOPS/mm$^2$) & 9.6       & 34.4     & 53.3     & -        & -     \\ \hline
\end{tabular}%
\label{table:sota-comparison}
\end{table*}

\section{Discussion}

The implemented ASIC design is optimized and tailored for gait analysis. As demonstrated in the results, it satisfies the application-specific requirements for both performance and accuracy. At the same time, to broaden its applicability, the processing architecture has been implemented in a fully parameterizable manner. Specifically, parameters, such as the number of input channels and the number of LSTM processing steps, are configured at startup by reading the associated data stored in EEPROM.

In addition, the implemented ASIC design supports the execution of the LSTM NN, including less than or equal to 20 LSTM cells. For cases requiring fewer than 20 LSTM cells, the parameters associated with unused cells can be set to zero, effectively disabling their contribution without adding computational overhead. Such a feature makes the architecture adaptable not only to various gait-analysis datasets but also to other low-frequency signal processing tasks, including ECG and related biomedical workloads. Importantly, any pre-trained neural network that conforms to the computational template described in this work can be directly deployed on the ASIC design.

Beyond the flexibility at the architectural level, the implemented ASIC design can also adopt various input sensors. Since the input interface and preprocessing stages are not tied to any specific sensor type, the design can ingest signals from various inertial units, pressure insoles, EMG electrodes, or other low-frequency biomedical sensors with minimal reconfiguration. By adjusting the input-channel parameters and updating the associated normalization or filtering coefficients at startup, the same hardware can be deployed across diverse data-acquisition setups. This sensor-agnostic design enables the chip to generalize beyond gait analysis and serve as a compact, low-power inference engine for broader wearable and physiological monitoring applications.

Finally, while this work focuses on a specific application, the multi-stage optimization workflow, spanning from application-level to physical synthesis, can be applied to designing edge AI accelerators. Given the computational demands of modern AI workloads and the typical resource constraints of edge devices, such a top-down co-design methodology provides a systematic means to harness the performance of advanced ML models while meeting stringent performance, power, and area requirements.
\section{Conclusions} 
\label{sec:conclusions}

In this paper, we presented the first cross-layer co-optimized LSTM accelerator for real-time gait analysis. Through a systematic approach, we minimized the bit-width of LSTM NN parameters and operations across various gait anomalies, while meeting accuracy and timing constraints. By software and hardware co-optimization, two architectural designs were selected for the physical design to indicate the trade-off between hardware complexity and accuracy in the LSTM accelerator. It was shown that the designed accelerators enable real-time gait analysis while meeting the timing requirement of the application, and are ready for the tape-out of an ASIC chip, which will be used in a real-world use-case. 


\section*{Acknowledgment}
This paper is supported in part the TEM‑TA138 project, co‑funded by the EU Structural Funds and the Estonian Ministry of Education and Research, EU Grant Project 101160182 “TAICHIP”, and the EU Grant 101194287 “NexTArc”.

\bibliographystyle{IEEEtran}
\bibliography{refs.bib}

\end{document}